\documentclass[aps,pre,twocolumn,longbibliography]{revtex4-1}

\usepackage{amssymb,amsmath}
\usepackage{graphicx}

\usepackage[latin1]{inputenc} 

\newcommand\beq{\begin{equation}}
\newcommand\eeq{\end{equation}}
\newcommand\beqa{\begin{eqnarray}}
\newcommand\eeqa{\end{eqnarray}}

\newcommand{\al}{\alpha}

\usepackage{xcolor}
\definecolor{darkgreen}{rgb}{0,0.6,0.0}

\begin{document}
\title{Energy nonequipartition  in a collisional model of a confined quasi-two-dimensional granular mixture}

\author{Ricardo Brito}
\email{brito@ucm.es}
\affiliation{Departamento de Estructura de la Materia, F\'isica T\'ermica y Electr\'onica and GISC, Universidad Complutense de Madrid, Spain}

\author{Rodrigo Soto}
\email{rsoto@dfi.uchile.cl}
\affiliation{Departamento de
F\'{\i}sica, Facultad de Ciencias F\'{\i}sicas y Matem\'aticas, Universidad de Chile, Santiago, Chile}

\author{Vicente Garz\'o}
\email{vicenteg@unex.es} \homepage{http://www.unex.es/eweb/fisteor/vicente/}
\affiliation{Departamento de
F\'{\i}sica and Instituto de Computaci\'on Cient\'{\i}fica Avanzada (ICCAEX), Universidad de Extremadura, E-06071 Badajoz, Spain}

\begin{abstract}

A collisional model of a confined quasi-two-dimensional granular mixture is considered to analyze homogeneous steady states. The model includes an effective mechanism to transfer the kinetic energy injected by vibration in the vertical direction to the horizontal degrees of freedom of grains. The set of Enskog kinetic equations for the velocity distribution functions of each component is derived first to analyze the homogeneous state. As in the one-component case, an exact scaling solution is found where the time dependence of the distribution functions occurs entirely through the granular temperature $T$. As expected, the kinetic partial temperatures $T_i$ of each component are different and hence, energy equipartition is broken down. In the steady state, explicit expressions for the temperature $T$ and the ratio of partial kinetic temperatures $T_i/T_j$ are obtained by considering Maxwellian distributions defined at the partial temperatures $T_i$. The (scaled) granular temperature and the temperature ratios are given in terms of the coefficients of restitution, the solid volume fraction, the (scaled) parameters of the collisional model, and the ratios of mass, concentration, and diameters. In the case of a binary mixture, the theoretical predictions are exhaustively compared with both direct simulation Monte Carlo and molecular dynamics simulations with a good agreement. The deviations are identified to be originated in the non-Gaussianity of the velocity distributions and on  microsegregation   patterns, which induce spatial correlations not captured in the Enskog theory.

 \end{abstract}

\pacs{05.20.Dd, 45.70.Mg, 51.10.+y, 47.50.+d}
\date{\today}
\maketitle

\section{Introduction}
\label{sec0}

Granular gases are paradigmatic examples of  intrinsically nonequilibrium  systems \cite{RevModPhys.68.1259}. A granular gas is composed of macroscopic particles (sizes varying from microns to millimeters or even larger \cite{poschel2001granular,Brilliantov9536}), where  kinetic energy of the colliding particles is transformed into internal degrees of freedom in form of vibration, deformation or microscopical fractures, but it is never returned back as kinetic energy. Such irreversible flux of energy leads to new and significantly different physics than in equilibrium statistical mechanics~\cite{RevModPhys.68.1259,andreotti2013granular}.

In equilibrium states, maximization of entropy leads to three conditions, called thermal, mechanical and chemical equilibrium conditions. The first one, thermal equilibrium, implies that every subsection of the system possess the same temperature. For instance, when applied to a  mixture of particles, equipartition implies that  different subspecies share the same temperature.  However, in non equilibrium states, like those studied here, maximization of entropy does not apply and energy equipartition may or may not be satisfied.  Violation of thermal equilibrium can show up in several ways. In fact, it has been reported that in a monocomponent granular system, different regions of the system can have different temperatures. In computer simulations, it was shown that a free evolving granular gas starting from a homogeneous initial condition forms dense and cold clusters, surrounded by a hot and dilute gas~\cite{GZ93}.

Energy equipartition is not verified either in mixtures of granular particles, which is the subject of the present paper. The first theoretical articles that described lack of energy equipartition in granular mixtures under homogeneous cooling state were reported  in the limits of tracer dynamics \cite{MP99} and Brownian motion \cite{BDS99},  and subsequently for general binary mixtures \cite{GD99b}. These theoretical results were confirmed by the Direct Simulation Monte Carlo (DSMC) method ~\cite{MG02,BT02} and by means of molecular dynamics (MD) simulations ~\cite{DHGD02,PMP02,BT02b,KT03,WJM03,BRM05,SUKSS06,KG14}.
In all cases, it was found that, despite the cooling rates for both types of particles being the same, they reach different temperatures.
These predictions motivated experiments, carried out by several researchers. In a three-dimensional vibrofluidized granular bed filled with particles of two sizes, tracked by positron emission tomography, it was found that the granular temperature of the larger particles was higher than that of the smaller diameter grains~\cite{WP02}. In a simpler setup consisting of two vertical glass plates with a  monolayer of granular particles inside, that is vertically vibrated, mixtures of  different types of particles (glass, aluminum, steel and brass) remain mixed, but reached different temperatures~\cite{FM02}. In the usual Brazil nut effect, also vibrated vertically, lack of energy equipartition induces a separation of hot/cold particles~\cite{BS09}.
Recent results \cite{BOB20}, have proved that the velocity distributions of granular mixtures (and therefore, the temperatures of each species or component) are quite universal, and do not depend on the particular mechanism of energy dissipation.
Finally, energy equipartition can be restored by a fine tuning of the particle material properties \cite{LVGS19}, so that the effect that the different sizes induce in the temperature differences can be compensated by concentration differences, for instance.

In the last years, there is a geometry that has attracted a great deal of attention in the study of granular systems (see Ref.~\cite{mujica2016dynamics} for a review). It is a confined geometry in the $z$-direction by means of two parallel plates
at a distance smaller than two particle diameters, so two particles cannot be on top of each other. The plates are large compared with the particle diameters, so it can be considered a quasi two-dimensional system in the $(x,y)$-plane. If the plates are made of glass, it is possible to camera-track position and velocities of the particles, allowing to obtain a detailed description of the particles dynamics in the plane. This makes possible to test and validate kinetic and hydrodynamic descriptions of the system. For systems with smooth plates, energy injection is carried out by the vertical vibration of the plates, mainly  transferred to the $z$-component of the particle velocities. By interparticle collisions, energy is then  transferred into the horizontal direction. As a conclusion,  vertical vibration acts as a thermostat in the horizontal plane and the inelastic collisions dissipate the injected energy.
However, when the plates have a rough surface, the dynamics is much more involved. In such case, they can directly inject energy to the horizontal degrees of freedom but the rotational degrees of freedom must be also considered in the theoretical description of the problem.

In the steady state, the temperature of the vertical degrees of freedom is higher than the horizontal ones, violating energy equipartition~\cite{rivas2011segregation,maynar2019understanding}.
Such setup, depending on density of particles and amplitude and/or frequency of vibration, can lead to a dynamical phase transition, where inhomogeneous patterns (clusters) are developed. Dense and cold clusters are formed, surrounded by hot, dilute gases violating  energy equipartition~\cite{PMEU04}. However, in a different range of parameters, the system remains homogeneous and a steady state is reached after a short term transient. We are interested in this regime.

Several effective two-dimensional models have been proposed to study this quasi two-dimensional geometry. The first simplification is to consider the granular particles as disks, as the motion in the third direction is very much confined (but not negligible). Energy injection can be modeled in several ways, depending if the plates as smooth or rough, that is, if horizontal momentum is conserved or not.
One approach, when momentum is not conserved,  is the stochastic thermostat, which adds a random force acting on every particle with zero correlation time, and amplitude related with the intensity of the kicks~\cite{WM96,peng1998steady,NE98,puglisi2012structure}.  Other models, which are appropriate when momentum is conserved, consider random restitution coefficients (larger or smaller than one) that lead to a homogeneous steady state~\cite{barrat2001inelastic,barrat2001heated,SGP15}.
In the present article we will consider the so-called $\Delta$-model also valid for smooth plates, where  the thermostat is a collisional one, so that energy is injected in every collision~\cite{brito2013hydrodynamic}. To be more precise, in a binary collision between particles of species $i$ and $j$, apart from the usual inelastic terms appearing in the collision rules characterized by the constant coefficient of restitution $\alpha_{ij}\leq 1$, an extra velocity $\Delta_{ij}$ is added to the normal component of the relative velocity of the two colliding spheres. Then, in a binary collision, the change in kinetic energy is composed by a dissipation energy term (proportional to $1-\alpha_{ij}^2$) plus two energy injection terms with  intensity depending on $\Delta_{ij}$. The origin of the $\Delta_{ij}$ term comes from the transfer of energy from the $z$-direction of vibration to the $(x,y)$-velocity plane when a collision between particles takes place. In real collisions, the energy transfer depends on both the component $z$ of the relative velocity of the colliding particles and the impact parameter in the $z$-direction. This means that the parameter $\Delta_{ij}$ is not a constant~\cite{maynar2019homogeneous}. On the other hand, for simplicity, we consider in this paper a \emph{constant} energy injection $\Delta_{ij}$, so that terms of the form $\pm \mu_{ij}\Delta_{ij} \widehat{{\boldsymbol {\sigma }}}$ are added to the velocities of the two colliding particles [$\widehat{{\boldsymbol {\sigma }}}$ being the unit vector joining the centers of the colliding spheres and $\mu_{ij}=m_i/(m_i+m_j)$, where $m_i$ and $m_j$ are the masses of the particles]. A study of the linear hydrodynamics for this model for a monocomponent system was done in the original paper \cite{brito2013hydrodynamic}. In a series of papers, Brey and coworkers  have extensively studied the $\Delta$-model in the past few years (see e.g.\ Ref.\ \cite{BBGM16} and references therein).

On the other hand, to the best of our knowledge, it must be remarked that up to now no comparison of the $\Delta$-model with experiments or simulations in confined geometry has been performed. However, it is expected that the model would be applicable in temporal scales where friction is weak, as for example in propagation of waves~\cite{Clerc08}.

How does the system behave when we consider granular mixtures of particles that evolve with the $\Delta_{ij}$-type of collisions?
Such is the goal of the present article: the study of a multicomponent mixture of $s$ different types of particles, having different masses ($m_i$), diameters ($\sigma_i$), coefficients of restitution ($\alpha_{ij}$), and inter-$\Delta$ parameters ($\Delta_{ij}$), where $i,j= 1,\ldots,s$. In particular, we focus on the stationary temperatures of the different components and the emergence of the violation of energy equipartition in the \emph{homogeneous} state.

The organization of the paper is as follows. In Sect.~\ref{sec1}, the extension of the $\Delta$-model to multicomponent granular mixtures is introduced. Then, the set of Enskog kinetic equations for the velocity distribution function of each component is displayed.
In Sect.~\ref{sec2}, we consider homogeneous solutions and derive evolution equations for the cooling rates and the temperatures of each component.  The steady state solutions to the above time-dependent equations are obtained in Sect.~\ref{sec3} where the partial cooling rates $\zeta_i$ associated with the partial temperatures $T_i$ are estimated by considering Maxwellian distributions defined at $T_i$. Implicit coupled equations for the temperatures $T_i$ are obtained by imposing that $\zeta_i=0$ in the steady state. As expected, in general the partial temperatures are different, violating energy equipartition. In addition, they exhibit a complex dependence on the parameter space of the system. The (approximate) theoretical results of the Enskog kinetic equation are confronted against computer simulations (carried out independently by both MD and the DSMC method) in Sect.~\ref{sec4} for different parameters of a binary mixture ($s=2$). This stringent test allows to gauge the validity of the different hypotheses (Maxwellian approximations, molecular chaos assumption and spatial correlations). The paper ends in Sect.~\ref{sec5} with a short summary and a discussion of the results obtained.

\section{Enskog kinetic equation. The collisional model for granular mixtures}
\label{sec1}

\subsection{Collisional model}

We consider a granular mixture of smooth inelastic hard spheres ($d=3$) or disks ($d=2$) of masses $m_i$ and diameters $\sigma_i$ ($i=1,\dots, s$). Here, we recall that $s$ means the number of components or species of the mixture. Let $(\mathbf{v}_1, \mathbf{v}_2)$ denote the pre-collisional velocities of two spherical particles of species $i$ and $j$, respectively. The collision rules for the post-collisional velocities $(\mathbf{v}_1',\mathbf{v}_2')$ are defined as
\beq
\label{1.1}
\mathbf{v}_1'=\mathbf{v}_1-\mu_{ji}\left(1+\alpha_{ij}\right)(\widehat{{\boldsymbol {\sigma }}}\cdot \mathbf{g})\widehat{{\boldsymbol {\sigma }}}-2\mu_{ji}\Delta_{ij} \widehat{{\boldsymbol {\sigma }}},
\eeq
\beq
\label{1.2}
{\bf v}_{2}'=\mathbf{v}_{2}+\mu_{ij}\left(1+\alpha_{ij}\right)(\widehat{{\boldsymbol {\sigma }}}\cdot \mathbf{g})\widehat{{\boldsymbol {\sigma }}}+2\mu_{ij}\Delta_{ij} \widehat{{\boldsymbol {\sigma }}},
\eeq
where we recall that $\mu_{ij}=m_i/(m_i+m_j)$, $\mathbf{g}=\mathbf{v}_1-\mathbf{v}_2$ is the relative velocity between the two colliding spheres, $\widehat{{\boldsymbol {\sigma}}}$ is the unit collision vector joining the centers of the two colliding spheres and pointing from particle 1 of component $i$ to particle 2 of component $j$. Particles are approaching if $\widehat{{\boldsymbol {\sigma}}}\cdot \mathbf{g}>0$. In Eqs.\ \eqref{1.1} and \eqref{1.2}, $0<\al_{ij}\leq 1$ is the (constant) coefficient of normal restitution for collisions $i$-$j$, and $\Delta_{ij}$ is an extra velocity added to the relative motion, with $\alpha_{ji}=\alpha_{ij}$ and $\Delta_{ji}=\Delta_{ij}$. This extra velocity points outward in the normal direction $\widehat{\boldsymbol {\sigma}}$, as required by the conservation of angular momentum \cite{L04bis}. The relative velocity after collision is
\beq
\label{1.3}
\mathbf{g}'=\mathbf{v}_1'-\mathbf{v}_2'=\mathbf{g}-(1+\al_{ij})(\widehat{{\boldsymbol {\sigma}}}\cdot \mathbf{g})
\widehat{\boldsymbol {\sigma}}-2\Delta_{ij} \widehat{{\boldsymbol {\sigma }}},
\eeq
so that the normal component of $\mathbf{g}'$ verifies the identity
\beq
\label{1.4}
(\widehat{{\boldsymbol {\sigma}}}\cdot \mathbf{g}')=-\al_{ij} (\widehat{{\boldsymbol {\sigma}}}\cdot \mathbf{g})-2\Delta_{ij}.
\eeq
As said in the Introduction, the model defined by the collisional rules \eqref{1.1}--\eqref{1.2} will be referred in this paper as the $\Delta$-collisional model.

With the set of collision rules \eqref{1.1} and \eqref{1.2}, momentum is conserved but energy is not. The change in kinetic energy upon collision is
\beqa
\label{1.5}
\Delta E_{ij}&\equiv& \frac{m_i}{2}\left(v_1^{'2}-v_1^2\right)+\frac{m_j}{2}\left(v_2^{'2}-v_2^2\right)\nonumber\\
&=&2m_{ij}\left[\Delta_{ij}^2+\al_{ij} \Delta_{ij} (\widehat{{\boldsymbol {\sigma}}}\cdot \mathbf{g})-\frac{1-\al_{ij}^2}{4}(\widehat{{\boldsymbol {\sigma}}}\cdot \mathbf{g})^2\right],\nonumber\\
\eeqa
where $m_{ij}\equiv m_im_j/(m_i+m_j)$ is the reduced mass. The right-hand side of Eq.\ \eqref{1.5} vanishes for elastic collisions ($\al_{ij}=1$) and $\Delta_{ij}=0$. Moreover, it is quite apparent from Eq.\ \eqref{1.5} that the kinetic energy can be gained or lost in a collision depending on whether $\widehat{{\boldsymbol {\sigma}}}\cdot \mathbf{g}$ is smaller than or larger than $2\Delta_{ij} /(1-\al_{ij})$.

The collision rules for the so-called \emph{restituting} collisions $\left(\mathbf{v}_1'',\mathbf{v}_2''\right)\to \left(\mathbf{v}_1,\mathbf{v}_2\right)$ with the same collision vector $\widehat{{\boldsymbol {\sigma }}}$ are defined as
\beq
\label{1.6}
\mathbf{v}_1''=\mathbf{v}_1-\mu_{ji}\left(1+\alpha_{ij}^{-1}\right)(\widehat{{\boldsymbol {\sigma }}}\cdot \mathbf{g})\widehat{{\boldsymbol {\sigma }}}-2\mu_{ji}\Delta_{ij}\al_{ij}^{-1} \widehat{{\boldsymbol {\sigma }}},
\eeq
\beq
\label{1.7}
\mathbf{v}_2''=\mathbf{v}_2+\mu_{ij}\left(1+\alpha_{ij}^{-1}\right)(\widehat{{\boldsymbol {\sigma }}}\cdot \mathbf{g})\widehat{{\boldsymbol {\sigma }}}+2\mu_{ij}\Delta_{ij}\al_{ij}^{-1} \widehat{{\boldsymbol {\sigma }}}.
\eeq

In the case of a direct collision, the volume transformation in velocity space is given by $d \mathbf{v}_1' d \mathbf{v}_2'=\al_{ij} d \mathbf{v}_1 d\mathbf{v}_2$, while in the case of an inverse (or restitution collision) one has $d\mathbf{v}_1'' d\mathbf{v}_2''=\al_{ij}^{-1} d\mathbf{v}_1 d\mathbf{v}_2$. All these relations will be employed later for evaluating some collision integrals.

\subsection{Enskog kinetic equation}

The extension of the Enskog kinetic equation for the above collision model to granular mixtures can be easily done by considering its version for its monocomponent case \cite{brito2013hydrodynamic,BGMB13,GBS18}. For an $s$-component mixture, the relevant information at a kinetic level  on the state of the system is given by the knowledge of the one-particle velocity distribution functions $f_i(\mathbf{r}, \mathbf{v}, t)$. For moderate densities and in the absence of external forces, the distributions $f_i(\mathbf{r}, \mathbf{v}, t)$ of the collisional model obey the set of coupled Enskog kinetic equations
\beq
\label{1.8}
\frac{\partial}{\partial t}f_i(\mathbf{r}, \mathbf{v};t)+\mathbf{v}\cdot \nabla f_i(\mathbf{r},\mathbf{v};t)=\sum_{j=1}^s\; J_{ij}[\mathbf{r},\mathbf{v}|f_i,f_j].
\eeq
The Enskog collision operators $J_{ij}$ of the model read
\beqa
\label{1.9a}
& & J_{ij}[\mathbf{r},\mathbf{v}_1|f_i,f_j]= \nonumber\\
& & \sigma_{ij}^{d-1}\int d{\bf v}_{2}\int  d\widehat{\boldsymbol{\sigma}}
\Theta (-\widehat{{\boldsymbol {\sigma }}}\cdot {\bf g}-2\Delta_{ij})\nonumber\\
& & \times
(-\widehat{\boldsymbol {\sigma }}\cdot {\bf g}-2\Delta_{ij})
\al_{ij}^{-2}\chi_{ij}(\mathbf{r},\mathbf{r}+\boldsymbol{\sigma}_{ij}) f_i(\mathbf{r},\mathbf{v}_1'';t)\nonumber\\
& & \times
f_j(\mathbf{r}+\boldsymbol{\sigma}_{ij},\mathbf{v}_2'';t)
-\sigma_{ij}^{d-1}\int\ d{\bf v}_{2}\int d\widehat{\boldsymbol{\sigma}}
\Theta (\widehat{{\boldsymbol {\sigma }}}\cdot {\bf g})\nonumber\\
& & \times
(\widehat{\boldsymbol {\sigma }}\cdot {\bf g})
\chi_{ij}(\mathbf{r},\mathbf{r}+\boldsymbol{\sigma}_{ij}) f_i(\mathbf{r},\mathbf{v}_1;t)
f_j(\mathbf{r}+\boldsymbol{\sigma}_{ij},\mathbf{v}_2;t),\nonumber\\
\eeqa
where $\Theta(x)$ is the Heaviside step function, $\boldsymbol{\sigma}_{ij}=\sigma_{ij}\widehat{\boldsymbol{\sigma}}$, and $\sigma_{ij}=(\sigma_i+\sigma_j)/2$.

As it happens in the Boltzmann equation, the expression \eqref{1.9a} of the Enskog collision operator neglects velocity correlations among the particles that are about to collide (molecular chaos hypothesis), and consequently the two-body distribution function factorizes into the product of the one-particle distribution functions. On the other hand, in contrast to the Boltzmann equation, it takes into account the spatial correlations [through the pair correlation functions at contact $\chi_{ij}(\mathbf{r},\mathbf{r}\pm\boldsymbol{\sigma}_{ij})$] between the colliding pairs as well as the variation of the distribution functions over distances of the order of the diameters of spheres \cite{G19}. Moreover, although the system considered is two-dimensional, our calculations will be carried out for an arbitrary number of dimensions $d$.

An important property of the Enskog collision operator $J_{ij}[f_i,f_j]$ is that the production term due to collisions
\beq
\label{1.9c}
\sigma_{\psi_i}\equiv \int\; d \mathbf{v}_1\; \psi_i(\mathbf{v}_1) J_{ij}[\mathbf{r},\mathbf{v}_1|f,f],
\eeq
can be expressed in a more convenient form than in \eqref{1.9c} by using some properties of the Enskog collision operator. It is given by \cite{BGMB13,SRB14}
\beqa
\label{1.9b}
\sigma_{\psi_i}&\equiv& \int\; d \mathbf{v}_1\; \psi_i(\mathbf{v}_1) J_{ij}[\mathbf{r},\mathbf{v}_1|f,f]\nonumber\\
&=&\sigma_{ij}^{d-1}\int d \, \mathbf{v}_1\int\ d {\bf v}_{2}\int d \widehat{\boldsymbol{\sigma}}\,
\Theta (\widehat{{\boldsymbol {\sigma }}}\cdot {\bf g})(\widehat{\boldsymbol {\sigma }}\cdot {\bf g})\nonumber\\
& & \times  \chi_{ij}(\mathbf{r},\mathbf{r}+\boldsymbol{\sigma}_{ij}) f_i(\mathbf{r},\mathbf{v}_1;t)
f_j(\mathbf{r}+\boldsymbol{\sigma}_{ij},\mathbf{v}_2;t)\nonumber\\
& & \times
\left[\psi_i(\mathbf{v}_1')-\psi_i(\mathbf{v}_1)\right].
\eeqa
The property \eqref{1.9b} is identical to the one obtained in the conventional inelastic hard sphere (IHS) model ($\Delta_{ij}=0$) \cite{BP04,G19}, except that $\mathbf{v}_1'$ is defined here by Eq.\ \eqref{1.1}.

\section{Homogeneous time-dependent state}
\label{sec2}

Let us consider spatially homogeneous isotropic states. In this case, Eq.\ \eqref{1.8} becomes
\beq
\label{2.1}
\frac{\partial}{\partial t}f_i(\mathbf{v};t)=\sum_{j=1}^N\; J_{ij}[\mathbf{v}|f_i,f_j],
\eeq
where $J_{ij}[f_i,f_j]$ is defined by Eq.\ \eqref{1.9a} with the replacements $\chi_{ij}(\mathbf{r},\mathbf{r}\pm\boldsymbol{\sigma}_{ij})\to \chi_{ij}$ and $f_j(\mathbf{r}\pm\boldsymbol{\sigma}_{ij},\mathbf{v};t)\to f_j(\mathbf{v};t)$. This means that the Enskog collision operator $J_{ij}=\chi_{ij}J_{ij}^\text{B}$, where $J_{ij}^\text{B}$ is the Boltzmann collision operator and $\chi_{ij}$ is the (homogeneous) pair correlation function at contact for collisions $i$-$j$. For practical purposes, and to agree with the equilibrium limit for elastic collisions ($\al_{ij}=1$ and $\Delta_{ij}=0$), $\chi_{ij}$ is usually taken to be the \emph{equilibrium} pair correlation function.

The collision operators conserve the particle number of each component and the total momentum:
\beq
\label{2.2}
\int d \mathbf{v}\; J_{ij}[\mathbf{v}|f_i,f_j]=0,
\eeq
\beq
\label{2.2.1}
\sum_{i=1}^s\sum_{j=1}^s \int d \mathbf{v}\; m_i \mathbf{v} J_{ij}[\mathbf{v}|f_i,f_j]=0.
\eeq
Nevertheless, in accordance with Eq.\ \eqref{1.5}, the total kinetic energy is not conserved in collisions and hence, the operators $J_{ij}$ verify the constraint
\begin{equation}
\sum_{i=1}^s\sum_{j=1}^s\int d {\bf v}\; m_{i}v^{2}J_{ij}[{\bf v}|f_{i},f_{j}]=-dnT\zeta \;,
\label{2.3}
\end{equation}
where $\zeta$ is the total cooling rate. It gives the rate of energy lost due to collisions among all components. In Eq.\ \eqref{2.3}, $n=\sum_i n_i$ is the total number density,
\beq
\label{2.4}
n_i= \int d \mathbf{v}\; f_i(\mathbf{v})
\eeq
is the number density of the component $i$, and
\beq
\label{2.5}
T=\frac{1}{d n}\sum_{i=1}^s\; \int d {\bf v}\; m_{i}v^{2} f_i(\mathbf{v})
\eeq
is the (total) granular temperature.

Apart from the granular temperature $T$, it is also convenient to introduce the partial temperatures $T_i$ defined as
\beq
\label{2.6}
T_i=\frac{1}{d n_i} \int d {\bf v}\; m_{i}v^{2} f_i(\mathbf{v}).
\eeq
The temperatures $T_i$ provide a measure of the mean kinetic energy of the component $i$. Comparison between Eqs.\ \eqref{2.5} and \eqref{2.6} yields the identity
\begin{equation}
T=\sum_{i=1}^s\;x_iT_i,
\label{2.7}
\end{equation}
where $x_i=n_i/n$ is the concentration or mole fraction of the component $i$. The ``cooling rates'' associated with the partial
temperatures can be defined by
\begin{equation}
\label{2.8}
\zeta_i=\sum_{j=1}^s \zeta_{ij}=-
\frac{1}{dn_iT_i}\sum_{j=1}^s\int d{\bf v}m_iv^{2}J_{ij}[{\bf v}|f_{i},f_{j}],
\end{equation}
where the second equality defines the quantities $\zeta_{ij}$. From Eqs.\ \eqref{2.3} and \eqref{2.8}, one easily gets the relation
\beq
\label{2.9a}
\zeta=\sum_{i=1}^s\; x_i \gamma_i \zeta_i,
\eeq
where $\gamma_i\equiv T_i/T$ is the temperature ratio of component $i$. The deviation of $\gamma_i$ from 1 provides a measure of the departure from energy equipartition (i.e., when $T_i=T$ for any component $i$).

The time evolution of $T_i$ and $T$ follow directly from the Enskog equation \eqref{2.1} and the definitions \eqref{2.3} and \eqref{2.8}:
\begin{equation}
\label{2.9b}
\zeta_i=-\frac{\partial}{\partial t}\ln T_i\;,\quad
\zeta=-\frac{\partial}{\partial t}\ln T.
\end{equation}
It is also convenient to write the time evolution of the temperature ratios $\gamma_i(t)=T_i(t)/T(t)$. Its time evolution can be easily obtained from Eq.\ \eqref{2.9b} as
\begin{equation}
\label{2.10}
\frac{\partial}{\partial t}\ln \gamma_{i}=\left(\zeta-\zeta_i\right).
\end{equation}

As usual, after a transient regime, in the same way as the monocomponent $\Delta$-model case~\cite{BMGB14} it is \emph{assumed} that there is  a special solution ({\em normal} or hydrodynamic solution \cite{CC70}) in which all the time dependence of the distribution functions occurs through the global temperature of the mixture $T(t)$. Assuming a normal form for multicomponent systems, it follows from dimensional analysis that $f_i(\mathbf{v},t)$ must be of the form
\begin{equation}
\label{2.11}
f_i(\mathbf{v},t)=n_iv_{\text{th}}^{-d}(t)\varphi_i\left(\mathbf{c}, \Delta_{\ell j}^*\right),\quad \ell, j=1,\ldots, s,
\end{equation}
where $\mathbf{c}\equiv \mathbf{v}/v_{\text{th}}$, $v_{\text{th}}(t)=\sqrt{2T(t)/\overline{m}}$ being a thermal velocity
defined in terms of the temperature of the mixture $T(t)$. In addition, $\overline{m}=\sum_i m_i/s$ and $\Delta_{ij}^*(t)\equiv \Delta_{ij}/v_{\text{th}}(t)$. The consistency of the assumption \eqref{2.11} will be verified a posteriori when we compare the theoretical predictions (which are based on the normal solution \eqref{2.11}) for the global temperature and the partial temperatures against computer simulations. As we will show later, the good agreement found between theory and simulations (specially in the low-density regime) confirms the reliability of the hypothesis \eqref{2.11}.

Contrary to the conventional IHS model \cite{NE98,GD99b}, the scaling distribution $\varphi_i$ depends on $T$ not only through the dimensionless velocity $\mathbf{c}$ but also through the dimensionless characteristic velocities $\Delta_{ij}^*\propto T(t)^{-1/2}$. Consequently,
\beq
\label{2.12}
T\frac{\partial f_i}{\partial T}=-\frac{1}{2}\frac{\partial }{\partial \mathbf{v}}\cdot \left(\mathbf{v}f_i\right)
-\frac{1}{2}\sum_{\ell, j=1}^N\; \Delta_{\ell j}^*\frac{\partial f_i}{\partial \Delta_{\ell j}^*}.
\eeq
In terms of dimensionless quantities, the Enskog equation \eqref{2.1} for the scaled distribution $\varphi_i$ can be written as
\beq
\label{2.13}
\frac{1}{2}\zeta^*\left(\frac{\partial}{\partial \mathbf{c}}\cdot \left(\mathbf{c}\varphi_i\right)
+\sum_{\ell, j=1}^N\; \Delta_{\ell j}^*\frac{\partial \varphi_i}{\partial \Delta_{\ell j}^*}\right)=\sum_{j=1}^s\; J_{ij}^*[\mathbf{c}|\varphi_i,\varphi_j],
\eeq
where
\beqa
\label{2.14}
& & J_{ij}^*[\mathbf{c}|\varphi_i,\varphi_j]\equiv \frac{v_\text{th}^d}{n_i \nu}J_{ij}[\mathbf{v}|f_i,f_j]=x_j\chi_{ij}\left(\frac{\sigma_{ij}}{\overline{\sigma}}\right)^{d-1}\nonumber\\
& &\times
\int d \mathbf{c}_{2}\int d \widehat{\boldsymbol {\sigma }}\,
\Theta (-\widehat{{\boldsymbol {\sigma }}}\cdot {\bf g}^*-2\Delta_{ij}^*)
(-\widehat{\boldsymbol {\sigma }}\cdot {\bf g}^*-2\Delta_{ij}^*)
\alpha_{ij}^{-2}\nonumber\\
& & \times \varphi_{i}(\mathbf{c}_1'')\varphi_{j}(\mathbf{c}_2'')-x_j\chi_{ij}\left(\frac{\sigma_{ij}}{\overline{\sigma}}\right)^{d-1}
\int d \mathbf{c}_{2}\int d \widehat{\boldsymbol {\sigma }}
\nonumber\\
& & \times
\Theta (\widehat{{\boldsymbol {\sigma }}}\cdot {\bf g}^*)
(\widehat{\boldsymbol {\sigma }}\cdot {\bf g}^*)\varphi_{i}(\mathbf{c}_1)
\varphi_{j}(\mathbf{c}_2)\;.
\eeqa
Here, $\nu=n\overline{\sigma}^{d-1}v_\text{th}$ is an effective collision frequency, $\overline{\sigma}=\sum_i \sigma_i/s$, and
$\mathbf{g}^*\equiv \mathbf{g}/v_\text{th}$. From Eq.\ \eqref{2.13}, one easily gets the dimensionless version of Eq.\ \eqref{2.10} as
\beq
\label{2.15}
\frac{1}{2}\zeta^* \sum_{\ell, j=1}^s\; \Delta_{\ell j}^*\frac{\partial \gamma_i}{\partial \Delta_{\ell j}^*}=
\gamma_i\left(\zeta^*-\zeta_i^*\right),
\eeq
where $\zeta^*=\zeta/\nu$ and
\begin{equation}
\label{2.16}
\zeta_i^*=\frac{\zeta_i}{\nu}=\sum_{j=1}^s \zeta_{ij}^*=-\frac{2}{d}\theta_i \sum_{j=1}^s \int\;
d {\bf c}\; c^{2}J_{ij}^*[\varphi_i,\varphi_j],
\end{equation}
where
\begin{equation}
\label{2.17}
\theta_i=\frac{m_i}{\overline{m}\gamma_i}.
\end{equation}
Note that the temperature ratios $\gamma_i$ are subjected to the constraint \eqref{2.7}. As a consequence, there are $s-1$ independent temperature ratios.

In summary, the homogeneous time-dependent solution is defined by the $s$ coupled equations \eqref{2.13} and the $s-1$ equations \eqref{2.15}. These $2s-1$ equations must be solved to obtain the $s$ scaling distributions $\varphi_i$ along with the $s-1$ temperature ratios $\gamma_i$. Approximate expressions for all the above unknowns are obtained by considering the simplest approximation: Maxwellian or Gaussian distributions at $T_i$. This approximate solution will be worked out in Sec.\ \ref{sec3}.

\section{Steady state solution. Maxwellian approximation}
\label{sec3}

We consider the steady state solution. In this case, for given values of $\Delta_{ij}^*$, $\partial_t T_i=0$ and so the partial cooling rates vanish in accordance with
Eq.\ \eqref{2.9b}:
\beq
\label{3.0}
\zeta^*=\zeta_1^*=\zeta_2^*=\cdots=\zeta_s^*=0.
\eeq
The determination of $\zeta_i^*$ requires the knowledge of the scaling distributions $\varphi_i$, whose exact form is not known to date. As in the conventional IHS model \cite{GD99b}, the distributions $\varphi_i$ can be expanded in a series of Sonine polynomials, the coefficients (cumulants) of the series being the corresponding velocity moments of $\varphi_i$. Usually, the first two terms are retained in the series expansion; the second term is related to the \emph{kurtosis} and measures the deviation of $\varphi_i$ from its Maxwellian form. Given that the above cumulants are small for conditions  of practical interest, non-Gaussian corrections to $\varphi_i$ are neglected for practical purposes. In particular, for the conventional IHS model, the theoretical predictions for the temperature ratios obtained by considering Maxwellian forms for the distributions $f_i$ show an excellent agreement with computer simulation results \cite{MG02}, even for strong inelasticity and/or disparate values of the mass and diameter ratios. We expect that such a good agreement is also present in the $\Delta$-collision model.

Thus, to estimate the partial cooling rates $\zeta_i^*$, we take the simplest Maxwellian approximation
\beq
\label{3.1}
\varphi_i(\mathbf{c})\to \varphi_{i,\text{M}}=\pi^{-d/2} \theta_i^{d/2}\; e^{-\theta_i c^2},
\eeq
where $\theta_i$ is given by Eq.\ \eqref{2.17}. As in previous works on granular mixtures\cite{GD99b}, for the sake of convenience, $\varphi_{i,\text{M}}$ is defined in terms of the partial temperature $T_i$ instead of the (global) granular temperature $T$. In fact, its second velocity moment is
\beq
\label{3.1.1}
\int d\mathbf{c}\; c^2\; \varphi_{i,\text{M}}=\frac{d}{2}\frac{\overline{m}T_i}{m_iT}.
\eeq

With the Maxwellian approximation \eqref{3.1}, $\zeta_i^*$ can be computed and its expression is (see Appendix \ref{appA} for some technical details)
\begin{widetext}
\beqa
\label{3.3}
\zeta_{i}^*&=&\frac{4\pi^{(d-1)/2}}{d\Gamma\left(\frac{d}{2}\right)}\sum_{j=1}^s
x_j\chi_{ij}
\left(\frac{\sigma_{ij}}{\overline{\sigma}}\right)^{d-1}\mu_{ji}(1+\al_{ij})\theta_i^{-1/2}
\left(1+\theta_{ij}\right)^{1/2}
\left[1-\frac{1}{2}\mu_{ji}(1+\alpha_{ij})(1+\theta_{ij}) \right]\nonumber\\
& &-\frac{4\pi^{d/2}}{d\Gamma\left(\frac{d}{2}\right)}\sum_{j=1}^s x_j\chi_{ij}
\left(\frac{\sigma_{ij}}{\overline{\sigma}}\right)^{d-1}\mu_{ji}\Delta_{ij}^*\left[
\frac{2\mu_{ji}\Delta_{ij}^*}{\sqrt{\pi}}\theta_i^{1/2}\left(1+\theta_{ij}\right)^{1/2}
-1+\mu_{ji}(1+\al_{ij})\left(1+\theta_{ij}\right)\right],
\eeqa
\end{widetext}
where $\theta_{ij}=\theta_i/\theta_j=m_i\gamma_j/m_j\gamma_i$ gives the ratio between the mean-square
velocity of the particles of the component $j$ relative to that of the particles of the component $i$.

In the limit of mechanically equivalent particles ($m_i=m$, $\sigma_i=\sigma$, $\al_{ij}=\al$, and $\Delta_{ij}^*=\Delta^*$), $\gamma_i=1$, and Eq.\ \eqref{3.3} yields $\zeta_i^*=\zeta^*$, where
\beq
\label{3.4}
\zeta^*=\frac{\sqrt{2}\pi^{(d-1)/2}}{d\Gamma\left(\frac{d}{2}\right)}\chi \Big(1-\al^2-2\Delta^{*2}-\sqrt{2\pi}\al \Delta^*\Big).
\eeq
The expression \eqref{3.4} is consistent with the one obtained for monocomponent granular gases \cite{brito2013hydrodynamic,BMGB14}.

\subsection{Binary mixture}

The results derived so far apply for an $s$-component mixture. For the purposes of illustration, henceforth a binary mixture (s=2) will be considered. In this case, the relevant quantities are the steady (scaled) temperature $T^*$ (defined below) and the temperature ratio $T_1/T_2$. Both quantities can be determined from the conditions \eqref{3.0}, namely,
\beq
\label{3.5}
\zeta_1^*=0, \quad \zeta_2^*=0.
\eeq
The solution to Eqs.\ \eqref{3.5} with the expression \eqref{3.3} for the cooling rates provides $T^*$ and $T_1/T_2$ in terms of the parameter space of the problem: the mass ratio $m_1/m_2$, the ratio of diameters $\sigma_1/\sigma_2$, the concentration $x_1$, the volume fraction $\phi$, the coefficients of restitution $\al_{11}$, $\al_{22}$, and $\al_{12}$, and the dimensionless velocities $\Delta_{11}^*$, $\Delta_{22}^*$, and $\Delta_{12}^*$. The volume fraction $\phi$ is defined as
\beq
\label{3.7}
\phi=\sum_{i=1}^2\; \frac{\pi}{4}n_i \sigma_i^d.
\eeq
Moreover, we are essentially interested here in a two-dimensional ($d=2$) system. In this case, a good approximation for the pair distribution function is \cite{JM87}
\beq
\label{3.6}
\chi_{ij}=\frac{1}{1-\phi}+\frac{9}{16}\frac{\phi}{(1-\phi)^2}\frac{\sigma_i\sigma_j M_1}{\sigma_{ij}M_2},
\eeq
where $M_\ell=\sum_i x_i \sigma_i^\ell$ and $\phi$ is given by \eqref{3.7} with $d=2$.

To scale the granular temperature $T$, it is convenient to introduce the parameter $\Delta$ as
\beq
\label{3.8}
\Delta=\sqrt{\Delta_{11}^2+\Delta_{22}^2+\Delta_{12}^2}.
\eeq
Thus, the reduced (steady) temperature $T^*$ is defined as
\beq
\label{3.9}
T^*=\frac{T}{\overline{m}\Delta^2/2}.
\eeq


\section{Comparison with computer simulations}
\label{sec4}

\subsection{Temperatures}

The theoretical results derived in Sec.\ \ref{sec3} for $T^*$ and $T_1/T_2$ from the Enskog kinetic equation are essentially based on three different approximations: (i) the use of the simple Maxwellian approximation \eqref{3.1} to estimate the partial cooling rates $\zeta_i^*$; (ii) the absence of velocity correlations between the velocities of the particles that are about to collide in \eqref{1.9a}; and (iii) the approximation \eqref{3.6} for the pair distribution function at contact. Therefore, it is important to assess the reliability of these theoretical results by comparison with computer simulations.

We have carried out simulations in this paper by employing the standard simulation methods. The first one is the DSMC method \cite{B94} adapted to \emph{dilute} granular gases. Since the DSMC method solves numerically the set of Boltzmann equations, it also assumes molecular chaos hypothesis. However, this method goes beyond the Maxwellian approximation since it determines the \emph{exact} velocity distribution functions $f_i$. In this context, the comparison of DSMC results versus analytical results for very dilute systems ($\phi\to 0$) can be used to assess the accuracy of the Maxwellian approximation \eqref{3.1} for determining the cooling rates \eqref{3.3}. The second method is MD simulations. This method avoids any assumptions inherent in the kinetic theory description [molecular chaos and Eq.\ \eqref{3.6} for accounting the spatial correlations in the Enskog equation] and/or approximations (Maxwellian distributions) made for evaluating the partial temperatures. For both methods, we simulate directly the $\Delta$-model, i.e., particles move in two dimensions, with collisions described by the rules \eqref{1.1} and \eqref{1.2}.

\begin{figure}
{\includegraphics[width=0.8\columnwidth]{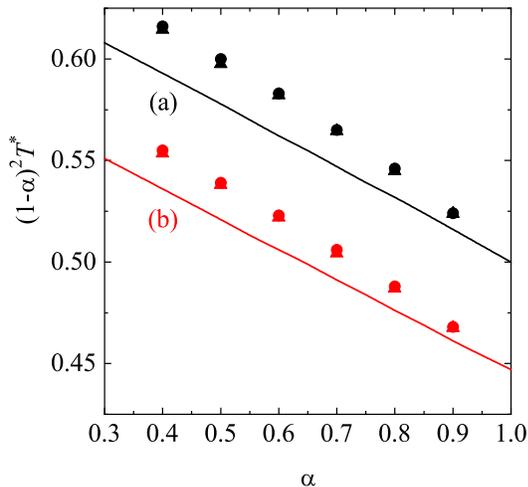}}
\caption{Case I. Scaled temperature $(1-\al)^2T^*$ versus $\alpha$ for $\sigma_1=2\sigma_2$, $\phi=0$, and two different values of the mass ratio $m_1/m_2$: $m_1/m_2=2$ (a) and $m_1/m_2=1/2$ (b). Symbols refer to DSMC results (circles) and MD simulations for $\phi=0.0016$ (triangles), and the lines correspond to the theoretical predictions derived from the Enskog equation.
\label{fig1}}
\end{figure}

\begin{figure}
{\includegraphics[width=0.8\columnwidth]{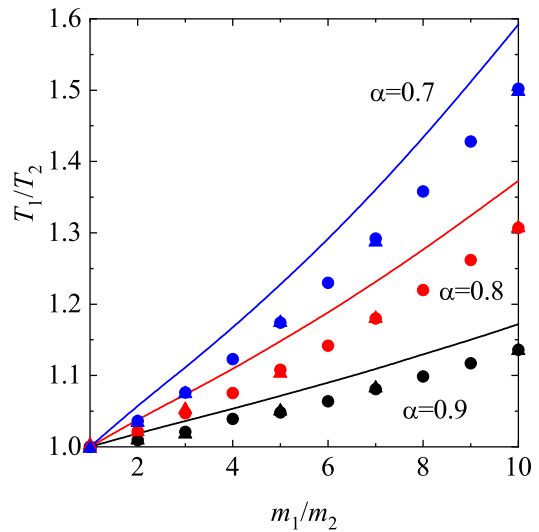}}
\caption{Case I. Temperature ratio $T_1/T_2$ versus the mass ratio $m_1/m_2$ for $\sigma_1=\sigma_2$, and three different values of the (common) coefficient of restitution $\alpha$: $\al=0.9$, 0.8 and 0.7. The lines refer to the Enskog theoretical results while the symbols correspond to the results obtained by numerically solving the Enskog equation by means of the DSMC method (circles) and by performing MD simulations for $\phi=0.0016$ (triangles). Note that the results obtained from DSMC and MD simulations are practically indistinguishable.
\label{fig2}}
\end{figure}

As in our previous papers on energy nonequipartition in granular mixtures
\cite{GD99b,MG02,DHGD02,GM03a,KG14,VLSG17,LVGS19,GGKG20}, given that the number of parameters involved in the problem is relatively high,
for the sake of simplicity we take a common coefficient of restitution $\al\equiv \al_{11}=\al_{22}=\al_{12}$, and an equimolar mixture $x_1=\frac{1}{2}$. For a monocomponent granular gas, the solution of Eq.~\eqref{3.4} gives that the reduced temperature scales with the coefficient of restitution as $T^*\propto1/(1-\alpha)^2$~\cite{brito2013hydrodynamic}. Hence, for the binary case, in order to compare results and obtain values of order one, we will present the scaled value $(1-\alpha)^2T^*$ instead of the scaled global temperature $T^*$.

Three different cases or systems are considered. In the following, we study each one of the cases separately.

\begin{center}
\textbf{CASE I}
\end{center}

We first analyze the usual case for binary mixtures, that is, when the components differ only in their masses and diameters. Hence, we assume here that $\Delta_{11}=\Delta_{22}=\Delta_{12}$ and we analyze the dependence of the scaled temperature $(1-\alpha)^2 T^*$ and the temperature ratio $T_1/T_2$ on the common coefficient of restitution $\alpha$, the parameters of the mass and diameter ratios ($m_1/m_2$ and $\sigma_1/\sigma_2$), and the solid volume fraction $\phi$.

The results for the global temperature, defined in Eq.~(\ref{3.9}), are shown in Fig.~\ref{fig1} as a function of  $\alpha$ for two different mass ratios and a very low density ($\phi=0.016$). Solid lines are the Enskog predictions, while symbols refer to DSMC and MD simulations. We observe first an excellent agreement between both methods of simulation, showing again the consistency of the DSMC method to numerically solve the Boltzmann equation. With respect to the comparison with the theoretical results, we see that the analytical results compare quite well with simulations, with deviations smaller than 4 or 5\% for large dissipation, i.e.\ for $\alpha=0.4$. Such deviations are due to non-Gaussian corrections to the velocity distribution functions. We have measured in DSMC the 4th cumulant, with the definition $\kappa_i=\langle v^4\rangle_i/2\langle v^2\rangle_i^2-1$ where
\beq
\label{4.1}
\langle v^\ell\rangle_i=\int d\mathbf{v}\; v^\ell\; f_i(v)/n_i.
\eeq
By definition, $\kappa_i$ vanishes when $f_i$ is approximated by its Maxwellian approximation \eqref{3.1}. For instance, for the simulation at $m_1/m_2=1/2$ and $\alpha=0.9$, we find that $\kappa_1=-0.046$ and $\kappa_2=-0.034$. These values increase with inelasticity, reaching at $\alpha=0.4$ the values of $\kappa_1=-0.13$ and $\kappa_2=-0.087$. Similar numbers are obtained for the other case ($m_1/m_2=2$) presented in Fig.~\ref{fig1}. Such values are the origin of the discrepancies observed in Fig.~\ref{fig1} between the Enskog theory and  DSMC simulations.

Figure~\ref{fig2} shows the lack of energy equipartition, as we plot the ratio $T_1/T_2$ for equal sized particles as a function of the mass ratio  in the case of small volume fraction. As in the conventional IHS model \cite{GD99b}, the temperature of the heavier particles is larger than that of the lighter ones since the temperature ratio $T_1/T_2$ increases steadily with the growing ratio of masses. Enskog results are plotted together with DSMC simulations (circles) and MD simulations  (triangles). As we see in this plot, both types of simulations agree again with great accuracy, while they separate from the Enskog prediction as the mass ratio grows. As before, the discrepancy is due to the use of the Maxwellian approximation \eqref{3.1} in the analytical calculation of $T_1/T_2$. In fact, the kurtosis of DSMC and MD are quite similar (for $\alpha=0.7$ and $m_1/m_2=10$ are $\kappa_1=-0.039, \kappa_2=-0.12$ for DSMC and  $\kappa_1=-0.034, \kappa_2=-0.12$ for MD).

\begin{figure}
{\includegraphics[width=0.8\columnwidth]{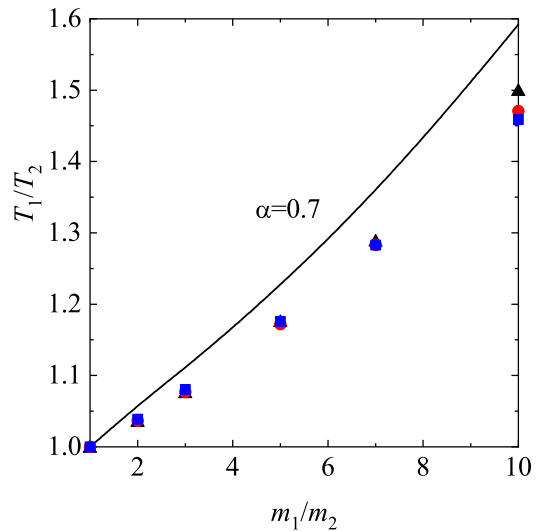}}
\caption{Case I. Temperature ratio $T_1/T_2$ versus the mass ratio $m_1/m_2$ for $\sigma_1=\sigma_2$, $\al=0.7$, and three different values of the volume fraction $\phi$: $\phi=0.0016$ (triangles), 0.1 (circles) and 0.2 (squares). Symbols refer to MD simulations and the line to the Enskog theoretical result.
\label{fig3}}
\end{figure}

To elucidate the role of density, we consider in what follows particles of equal size ($\sigma\equiv\sigma_1=\sigma_2$). This election has the advantage that, despite the Enskog results depend on $\phi$ through $\chi_{ij}(\phi)$, such dependence vanishes for the computation of $T^*$ and $T_1/T_2$ when considering particles of equal size and equimolar mixtures. Indeed, when $\sigma_1=\sigma_2$ and $x_1=\frac{1}{2}$, Eq.\ \eqref{3.6} yields $\chi_{11}=\chi_{22}=\chi_{12}$ and, hence, they factor in the equations \eqref{3.5}. As a result, the Enskog theory does not predict any dependence of the stationary temperatures on the global volume fraction in the situation above. This is a consequence of both the energy injection and  dissipation mechanisms being collisional. This cancellation takes place at the level of the Enskog theory but if there are position correlations not captured in the expression \eqref{3.6} for $\chi_{ij}$, density corrections may appear. The comparison with MD simulations carried out in Fig.\ \ref{fig3} tests this prediction, for which we make MD simulations at very low as well as moderate densities (up to $\phi=0.2$). The Enskog theory has shown to be quite accurate in the above range of densities for the IHS model of granular fluids (see the comparison with MD simulations in Refs.\ \cite{DHGD02,LBD02,MGAL06,LLC07,MDCPH11,MGH14} and with real experiments in \cite{YHCMW02,HYCMW04}). It is quite apparent from Fig.~\ref{fig3} that the density dependence in this case is rather weak, and mostly appears at high mass ratio, validating the Enskog hypothesis.

\begin{center}
\textbf{CASE II}
\end{center}

In Case II we consider that the two components are mechanically equivalent ($m_1=m_2$, $\sigma_1=\sigma_2$), except for the interparticle   energy injection, such that $\Delta_{11}=\Delta_{22}$, but $\Delta_{12}=\lambda \Delta_{11}$, with $\lambda \geq 0$. This case implies that collisions 1-2 are produced with a different $\Delta$ that 1-1 or 2-2 collisions.  For instance, if $\lambda=0$, 1-2 collisions do not gain energy, but are purely inelastic. However, if $\lambda>1$ the 1-2 collision gains more energy than the 1-1 or 2-2 collisions and the system heats up. Although this case is somehow artificial and difficult to implement in practice, its study is of interest. Similarly to undriven granular  mixtures with $\alpha_{11}=\alpha_{22}$, but $\alpha_{12}$ different~\cite{garzo2011thermal}, this case puts the theory to a stringent test because $T_1$ trivially equals $T_2$ by construction (both Enskog theory and computer simulations yield $T_1/T_2=1$). This makes easier the identification of the deviations from the theoretical predictions, as for example, our analysis on microsegregation in Section~\ref{sec.correlations}.

\begin{figure}
{\includegraphics[width=0.8\columnwidth]{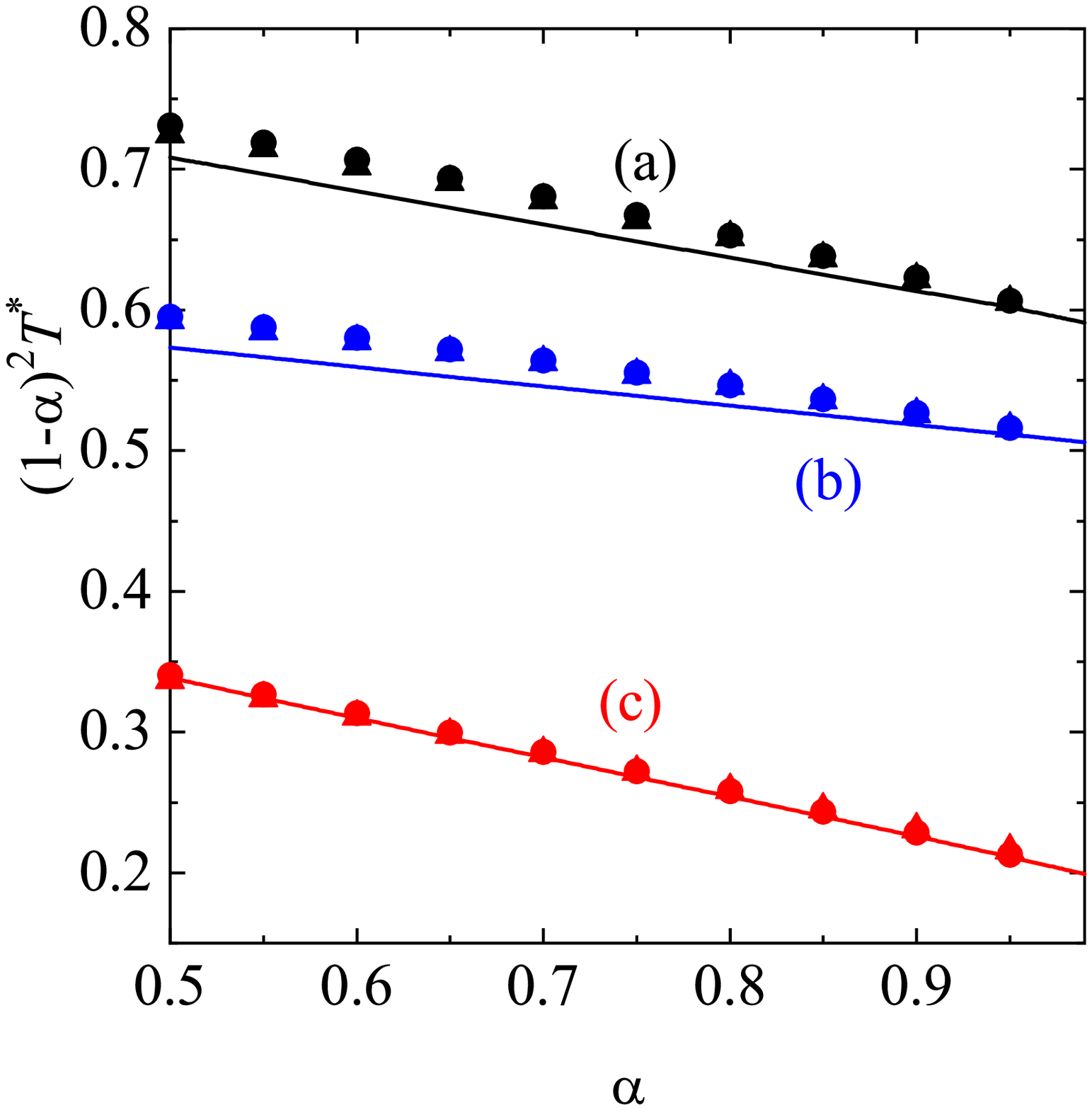}}
\caption{Case II. Plot of $(1-\al)^2T^*$ versus $\alpha$ for $m_1=m_2$, $\sigma_1=\sigma_2$, $\phi=0$, $\Delta_{11}=\Delta_{22}$, and $\Delta_{12}=\lambda \Delta_{11}$. Three different values of $\lambda$ have been considered: $\lambda=2$ (a), $\lambda=0.9$ (b), and $\lambda=0$ (c). Symbols refer to DSMC results (circles) and MD simulations (triangles) for $\phi=0.01$ while the lines correspond to the Enskog theoretical results.
\label{fig4}}
\end{figure}

The dependence of the scaled temperature $(1-\alpha)^2 T^*$  is studied for different values of $\lambda$ in Fig.~\ref{fig4} for a low density mixture. As mentioned before, in the case of $\lambda=0$ [case (c)] 1-2 collisions are purely dissipative, so one expects a lower temperature than the cases when $\lambda >0$. As $\lambda$ grows, the average temperature grows as well.
For this choice of parameters, both MD and DSMC simulations agree well with the Enskog theory. It is found for this case that the values of the kurtosis $\kappa_i$ are smaller than in Case I. It is worth noticing that the case of $\lambda=0$ has the smallest kurtosis, typically 4 times smaller that the cases (a) and (b). This means that its velocity distribution is close to a Maxwellian distribution and consequently the agreement with the  theory is excellent.

\begin{figure}
{\includegraphics[width=0.8\columnwidth]{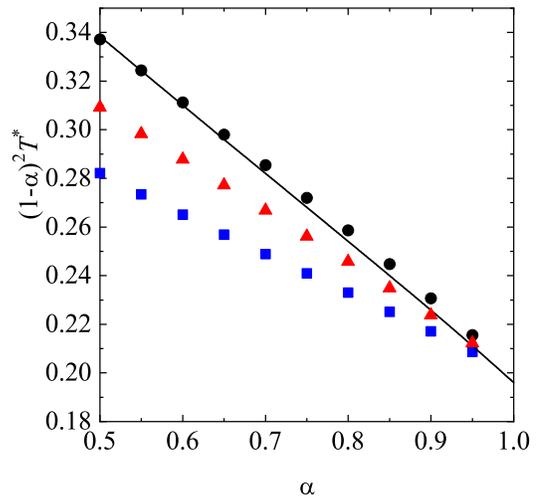}}
\caption{Case II. Plot of $(1-\al)^2T^*$ versus $\alpha$ for $m_1=m_2$, $\sigma_1=\sigma_2$, $\phi=0$, $\Delta_{11}=\Delta_{22}$, and $\Delta_{12}=\lambda \Delta_{11}$, with $\lambda=0$. Three values of the volume fraction $\phi$ are presented: $\phi=0.0016$ (triangles), 0.1 (circles) and 0.2 (squares). Symbols refer to MD simulations and the line to the Enskog theoretical result.
\label{fig4denso}}
\end{figure}

The effect of density is shown in Fig.~\ref{fig4denso}, where we have selected the case $\lambda=0$ for which the agreement between theory and simulations is excellent at low density. Important deviations of up to 20\% are obtained in the most inelastic case. These deviations are the effect of position and velocity correlations; while the first correlation is accounted for by the approximation \ \eqref{3.6} of $\chi_{ij}$, the second one is neglected in the Enskog theory. Subsection~\ref{sec.correlations} describes the origin of such correlations.

\begin{center}
\textbf{CASE III}
\end{center}

\begin{figure}
{\includegraphics[width=0.8\columnwidth]{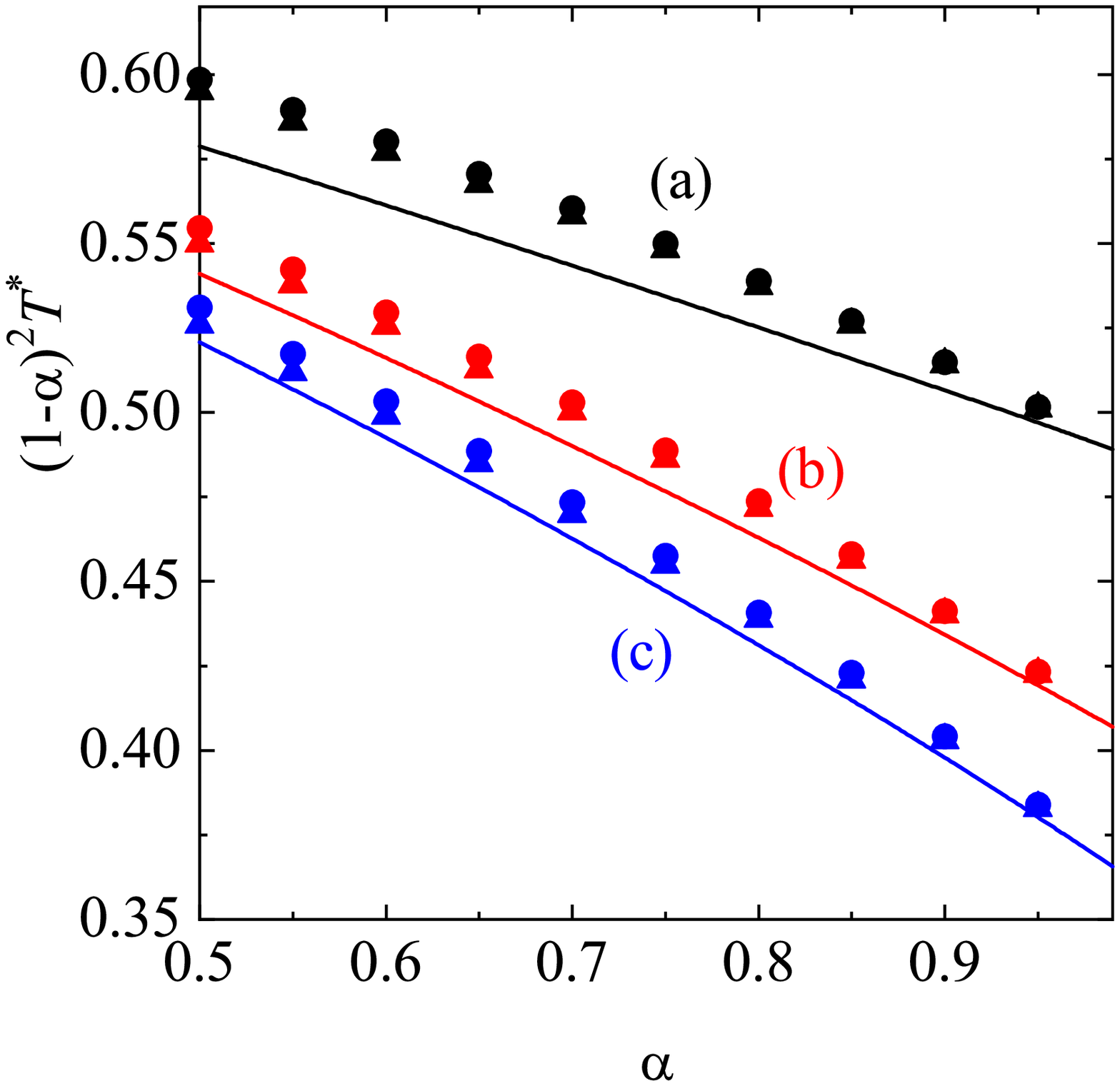}}
\caption{Case III. Plot of $(1-\al)^2T^*$ versus $\alpha$ for $m_1=m_2$, $\sigma_1=\sigma_2$, $\phi=0$, $\Delta_{22}=\lambda \Delta_{11}$, and  $\Delta_{12}=(\Delta_{11}+\Delta_{22})/2$. Three different values of $\lambda$ have been considered: $\lambda=2$ (a), $\lambda=5$ (b), and $\lambda=10$ (c). Symbols refer to DSMC results (circles) and MD simulations (triangles) for $\phi=0.01$ while the lines correspond to the Enskog theoretical results.
\label{fig5}}
\end{figure}

\begin{figure}
{\includegraphics[width=0.8\columnwidth]{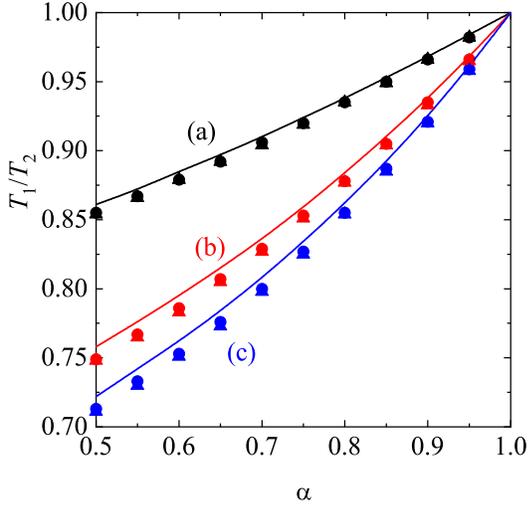}}
\caption{Case III. Plot of $T_1/T_2$ versus $\alpha$ for $m_1=m_2$, $\sigma_1=\sigma_2$, $\phi=0$, $\Delta_{22}=\lambda \Delta_{11}$, and  $\Delta_{12}=(\Delta_{11}+\Delta_{22})/2$. Three different values of $\lambda$ have been considered: $\lambda=2$ (a), $\lambda=5$ (b), and $\lambda=10$ (c). Symbols refer to DSMC results (circles) and MD simulations (triangles) for $\phi=0.01$ while the lines correspond to the Enskog theoretical results.
\label{fig6}}
\end{figure}

Here, we consider that the two components differ in the energy injection at collisions, such that $\Delta_{11} < \Delta_{22}$, and we take for simplicity $\Delta_{12}=(\Delta_{11}+\Delta_{22})/2$. Otherwise, the components of the mixture are mechanically equivalent ($m_1=m_2$, $\sigma_1=\sigma_2$). We analyze the dependence of $(1-\alpha)^2 T^*$ and $T_1/T_2$ on $\al$ for different values of the ratio $\Delta_{22}/\Delta_{11}$. In this case, the particles of type 1 reach a higher temperature than if they were alone, as collisions 1-2 inject more energy that collisions 1-1 (because $\Delta_{12}>\Delta_{11}$). On the contrary, particles of type 2 are cooler than if they were alone, due to the fact that $\Delta_{12}$ is smaller that $\Delta_{22}$.  In general, as Figs.~\ref{fig5} and \ref{fig6} show, the Enskog results agree well with DSMC and MD simulations at low density. It is apparent from Fig.~\ref{fig6} that the temperature ratio is clearly different from 1, showing again the lack of energy equipartition. In this case, the injection of energy of particles of type 1 is smaller than for particles of type 2, despite the interparticle collisions compensate that difference. The net balance is a ratio $T_1/T_2$ always smaller than 1.

\begin{figure}
{\includegraphics[width=0.8\columnwidth]{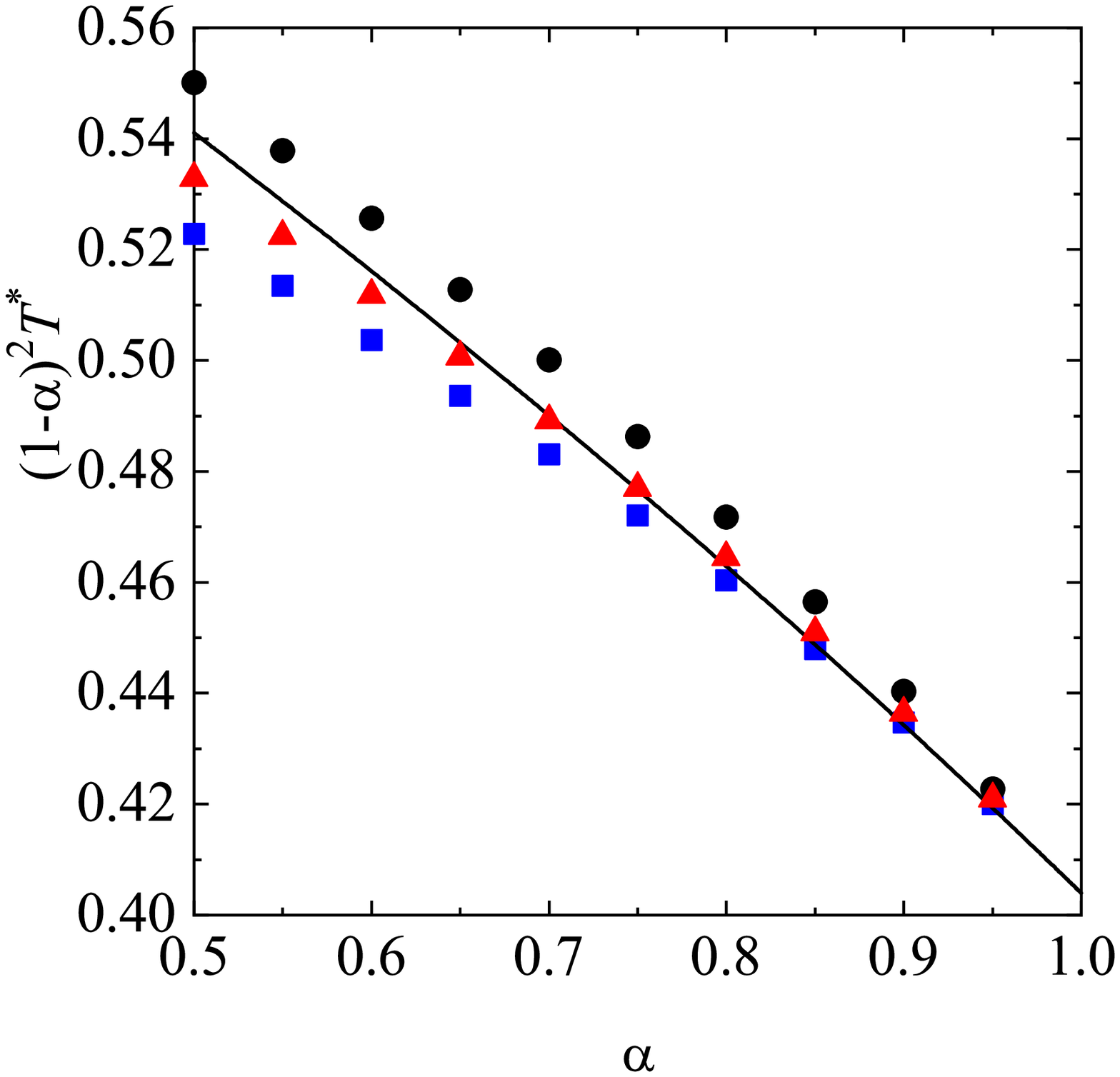}}
\caption{
Case III. Plot of $(1-\al)^2T^*$  versus $\alpha$ for $m_1=m_2$, $\sigma_1=\sigma_2$, $\lambda=5$, and three different values of density: $\phi=0.01$ (solid line and circles), $\phi=0.1$ (triangles), and $\phi=0.2$ (squares). Symbols refer to MD simulations and the line to the Enskog theoretical result.
\label{fig5dense}}
\end{figure}

\begin{figure}
{\includegraphics[width=0.8\columnwidth]{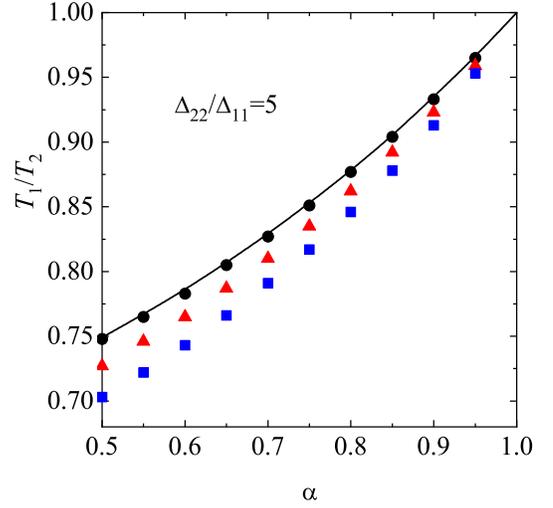}}
\caption{Case III. Plot of $T_1/T_2$ versus $\alpha$ for $m_1=m_2$, $\sigma_1=\sigma_2$, $\lambda=5$, and three different values of density: $\phi=0.01$ (solid line and circles), $\phi=0.1$ (triangles), and $\phi=0.2$ (squares). Symbols refer to MD simulations and the line to the Enskog theoretical result.
\label{fig7}}
\end{figure}

The Enskog results together with MD simulations for three different densities are plotted in Figs.\ \ref{fig5dense} and \ref{fig7}. We observe that the effect of density for Case III is larger than for Case I, but smaller than for Case II.  In this  case (see Fig.~\ref{fig5dense}) the effect of non Gaussianity and the density effects act in opposite directions. It is important to note again that kurtosis also shows a weak dependence with density and therefore the discrepancies with Enskog theory are mainly due to  correlations not captured in the theory, which are discussed in the next Subsection.

\subsection{Pair distribution functions} \label{sec.correlations}

\begin{figure}
{\includegraphics[width=0.9\columnwidth]{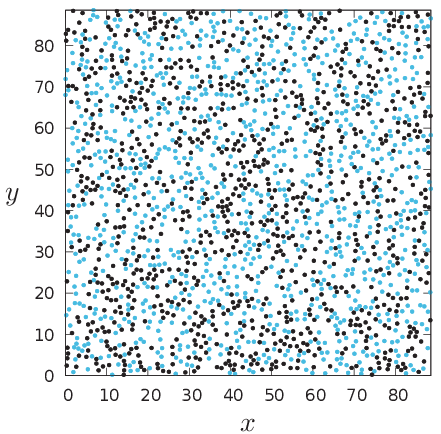}}
{\includegraphics[width=0.9\columnwidth]{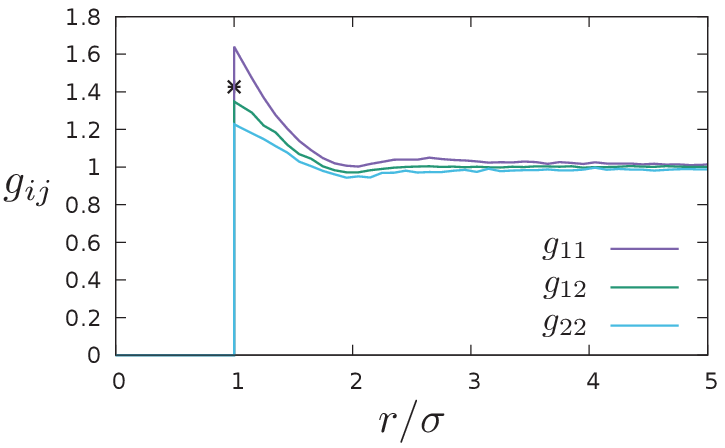}}
\caption{Molecular dynamic results for case III, with $\phi=0.2$, $\alpha=0.7$, and $\lambda\equiv \Delta_{22}/\Delta_{11}=5$. Top: snapshot of the system in the steady state. Black (blue) circles are particles of type 1 (2). Bottom: pair correlation functions $g_{ij}$ as a function of the interparticle distance $r/\sigma$. The (common) value of $\chi_{ij}$ predicted by Eq.\ \eqref{3.6} is represented by the asterisk.
\label{figconfigchi}}
\end{figure}

To study the development of spatial correlations, we analyze the spatial distribution of particles obtained in  MD simulations. Figure \ref{figconfigchi} presents the results for a dense simulation of case III, with a large contrast in the energy injection ($\lambda\equiv \Delta_{22}/\Delta_{11}=5$). According to Figs.~\ref{fig5dense} and \ref{fig7}, MD simulations show an important density dependence on $T^*$ and $T_1/T_2$. The top panel shows a snapshot, where it is evident that the system remains globally homogeneous, while there is a tendency for the less energetic particles (type 1, black particles in the figure) to aggregate because they separate at a smaller speed after collisions, compared with the case when a particle of type 2 is involved. This is a manifestation of \emph{microsegregation}. The displayed snapshot is in the stationary regime and the aggregates are dynamical, continuously forming and dissolving. No coarsening is observed.

The aggregation of type 1 particles makes 1-1 collisions more frequent than the mean-field estimation made by the Enskog theory. As these collisions are less energetic, $T_1$ is smaller than the prediction of the Enskog theory, resulting in the decrease of the global temperature  and the temperature ratio with density, as shown in Figs.~\ref{fig5dense} and \ref{fig7}.

In order to make this intuition more quantitative, we have measured the pair correlation functions $g_{ij}(r)$ in MD simulations. The pair correlation functions at contact are obtained as $\chi_{ij}=\lim_{r\to\sigma^+}g_{ij}(r)$. For inelastic collision rules, as for Eqs.~\eqref{1.1} and \eqref{1.2}, the pair correlation function at contact is discontinuous and depends on the  angle between the relative velocity $\mathbf{g}=\mathbf{v}_1-\mathbf{v}_2$ and the vector joining the center of the particles $\mathbf{r}=\mathbf{r}_1-\mathbf{r}_2$, distinguishing pairs that are about to collide from those that just collided~\cite{soto2001statistical}. The Enskog theory uses only the pair correlations of particles that are about to collide and, consequently, in the MD simulations we extract only the  correlation functions of pairs with $\mathbf{r}\cdot\mathbf{g}<0$, properly normalized such at large distances they approach unity. As advanced from the snapshot, $g_{11}>g_{12}>g_{22}$ at short distances, implying that Eq.\ \eqref{3.6} for $\chi_{ij}$ is not completely accurate [for $\phi=0.2$ and $x_1=\frac{1}{2}$, Eq.~\eqref{3.6} gives $\chi_{11}=\chi_{22}=\chi_{12} \simeq 1.43$]. As discussed before, MD simulations show a value of $\chi_{11}$ about a 20\% larger than that of Eq.~\eqref{3.6}, while $\chi_{22}$ is, however about  a 15\% smaller. These differences quantify the effect of the spatial correlations on the temperatures.

Deviations of $\chi_{ij}$ respect to their predicted theoretical ---equilibrium--- values, could be universal
in granular gases. For instance, it was observed in monocomponent randomly driven granular gases \cite{pagonabarraga2001randomly}, where it was found an increment in the pair distribution function $g(r)$, but the long range structure remains homogeneous. In the case of granular mixtures, we are only aware of one study \cite{BEGS08} where the pair distribution function was measured,  giving evidence of microsegregation, but in that case macroscopic segregation also occurred. The observed microsegregation results from recollision events, which are not included in the Enskog description. However, although these effects can be large in the pair correlation functions themselves (as shown in the bottom panel of Fig.~\ref{figconfigchi}), their impact on the Enskog predictions for $T^*$ and $T_1/T_2$ is not quite significant and only produces a discrepancy between theory and MD simulations that is not larger than 7\%. This excellent agreement justifies the use of the Enskog theory, which is much simpler than including recollision events in the theory.

The lower temperature obtained for Case II as increasing density (see Fig.~\ref{fig4denso}) is also a result of microsegregation. Indeed, for $\lambda=0$, when 1-2 particles collide, they separate slower than if the particles were of equal type, resulting in an effective attractive interaction between dissimilar particles. When recollisions take place as the density increases, the  frequency of 1-2 collisions is larger than the mean-field estimation made by the Enskog theory. As these collisions dissipate more energy, the global temperature is reduced. A similar reasoning can be used to show that for $\lambda>1$, the effect of density is also to decrease the global temperature (not shown).

\section{Discussion}
\label{sec5}

The present paper is focused on the study of the Enskog kinetic theory for mixtures of granular particles, evolving under the so-called $\Delta$-model. In particular, we are  interested here in obtaining the partial temperatures of the components of the mixture in a homogeneous state. As expected from previous works on IHS model \cite{GD99b,G19}, the lack of energy equipartition is also present in the $\Delta$-model as a consequence of the nonequilibrium nature of inelastic collisions.

The theoretical development starts with the extension of the $\Delta$-model to multicomponent granular mixtures. Then, following standard procedures, the set of Enskog kinetic equations for the velocity distribution functions of each component is provided where the nonlinear Enskog collision operators $J_{ij}[f_i,f_j]$ are given by Eq.\ \eqref{1.9a}. As happens in the conventional IHS model \cite{G19}, explicit results for the first few velocity moments of the distribution functions (granular hydrodynamics) can be derived thanks to the use of the property \eqref{1.9b} for the production term due to collisions. In general, the distribution functions contain a spatial dependence so hydrodynamic variables can be inhomogeneous, relaxing to homogeneous ones via the corresponding transport coefficients.

Before considering inhomogeneous states and in order to extract relevant results for the partial temperatures, we assume in this paper spatially homogeneous isotropic states where the hydrodynamic variables transform into homogeneous  variables. In the time-dependent problem, the distribution functions $f_i(\mathbf{v},t)$ of each component have the scaling form \eqref{2.11} with the time dependence provided by the (global) granular temperature $T(t)$, as required for a \emph{normal} or hydrodynamic solution. However, in contrast to the IHS model \cite{GD99b}, the dependence of the scaling functions $\varphi_i$ on $T$ is not only encoded through the scaled velocity $\mathbf{c}=\mathbf{v}/v_\text{th}$ ($v_\text{th}\propto \sqrt{T}$ being a thermal velocity) but also through the dimensionless parameters $\Delta_{ij}^*=\Delta_{ij}/v_\text{th}$. This type of scaling is common in driven granular mixtures \cite{GGKG20,KG13}.

Assuming the scaling \eqref{2.11}, we have derived evolution equations for the temperature ratios $\gamma_i(t)=T_i(t)/T(t)$ [see Eq.\ \eqref{2.15}]. These quantities measure the departure from the energy equipartition. The evolution equation for the temperatures leads, after a transient regime,  to stationary values obtained when the (reduced) partial cooling rates $\zeta_i^*$ vanish [see Eq.\ \eqref{2.9b}]. On the other hand, the computation of $\zeta_i^*$ requires the knowledge of the scaling distribution function $\varphi_i$. Since those functions are not exactly known, $\varphi_i$ is usually expanded in Sonine polynomials where only the first two terms are generally retained for practical purposes. However, based on the results derived in the IHS model for granular mixtures \cite{GD99b} where the non-Gaussian corrections to $\varphi_i$ are in general very small, we estimate here $\zeta_i^*$ by taking the simplest approximation for $\varphi_i$: the Maxwellian approximation \eqref{3.1}. It is important to remark that the Maxwellian distribution $\varphi_{i,\text{M}}$ is defined at the temperature of the component $i$, so that the corresponding Maxwellians for two components can be quite different due to the temperature differences.

For purposes of illustration, a binary mixture has been considered. In this case, the numerical solution to the conditions $\zeta_1^*=\zeta_2^*=0$ allows us to determine the (steady) temperature ratio $T_1/T_2$ as a function of the parameter space of the system: the mass ratio $m_1/m_2$, the diameter ratio $\sigma_1/\sigma_2$, the concentration $x_1$, the volume fraction $\phi$, the coefficients of restitution $\alpha_{ij}$, and the parameters $\Delta_{ij}^*$. Although our analytical results are approximate, we expect that they are not restricted to quasielastic systems and apply for arbitrary composition, mass ratio, particle diameter, and a wide range of
density (namely, an expected accuracy comparable to that of the Enskog theory for the conventional IHS model).

To assess the reliability of the (approximate) Enskog results, we have also performed MD and Monte Carlo (DSMC method) simulations of the same system. Since the parameter space for a  binary mixture is huge (10 independent parameters for evaluating $T_1/T_2$), for the sake of simplicity, we have considered here a (common) coefficient of restitution $\alpha$, an equimolar mixture $x_1=\frac{1}{2}$, and have selected some combination of parameters out of this huge parameter space. They are labelled Case I, II and III. An exhaustive comparison between Enskog theory and computer simulations have been performed separately for each one of the above Cases.

The main result is that the Enskog theory  is able to predict the temperatures of each component and capture the lack of energy equipartition with good accuracy. The validity of the Enskog prediction goes beyond the Boltzmann limit of low density, and compares quite well with simulations, both DSMC and MD. Each type of simulation allows us to identify the effect of the two approximations [Maxwellian approximation, use of the form \eqref{3.6} for $\chi_{ij}$, and absence of velocity correlations] carried out in the theoretical analysis. The DSMC method does not contain space correlations, but the velocity distribution functions are not Maxwellians since they are characterized by a kurtosis different from zero. We have observed that the impact
of the Gaussian approximation is always smaller than 10\% for the global temperature and the temperatures of the individual components. On the contrary, MD simulations avoids the above assumptions since contains the three effects: non-Gaussianity and spatial and velocity correlations. This allows us to demonstrate the lack of energy equipartition of the $\Delta$-model in a broader context. At high densities, the analysis of the spatial configurations and pair distribution functions shows microsegregation originated in the different dynamics of the components. Although these effects cannot be captured by the Enskog kinetic theory, they produce a small influence on the (steady) granular temperature and the temperature ratio, except for Case II (which can be considered as a somehow artificial case) for small coefficients of restitution. In conclusion, the present results give again support to the use of the Enskog equation for the description of granular flows across a wide range of densities, length scales, and inelasticity. Despite this success, the observed microsegregation opens the necessity of developing kinetic theories that go beyond the Enskog theory  but, as has been mentioned in several previous works \cite{G19}, no other theory with such generality exists yet.

In the extension to mixtures of the $\Delta$-model, we consider that the parameters $\Delta_{ij}$ could be chosen arbitrarily. However, in a confined three-dimensional system under vibration, they should be computed considering the full collisional geometry and the effect of the  vibrating plates. This a tremendous task with only partial known results up to now. In Ref.~\cite{maynar2019homogeneous}, the vertical-to-horizontal energy balance was obtained for the case of a monocomponent granular gas in a box with smooth elastic plates vibrating at infinite frequency but finite velocity $V_0$. By equating the energy balance [their Eq.~(31) with Eq.~\eqref{3.4} of this article], an approximate mapping to the $\Delta$-model can be made, obtaining the scaling $\Delta\sim V_0 [(H-\sigma)/\sigma]^3$, where $H$ is the plate separation. No direct scaling with the coefficient of restitution is possible as the functional form for the energy balance is not equal to that of  Eq.~\eqref{3.4}. Additionally, they showed that the vertical energy that is injected into the horizontal degrees of freedom increases with the time between grain-grain collisions as a result of successive collisions with the plates. Modeling this phenomenon as $\Delta$ increasing in time between collisions generates a gas-liquid phase transition~\cite{risso2018effective}. The muticomponent case is considerably more complex and futher research is needed to unveil the mapping between the parameters of the $\Delta$-model to those of the three dimensional system.

The results obtained in this paper opens up the possibility of deriving the Navier--Stokes hydrodynamics equations of the mixture with explicit forms of the corresponding transport coefficients. These coefficients can be obtained for instance by solving the set of Enskog equations for states with small spatial gradients by means of the Chapman--Enskog method adapted to inelastic collisions \cite{CC70,G19}. The reference state in this method is obtained from the \emph{local} version of the time-dependent homogeneous state defined by Eq.\ \eqref{2.11}, namely,
\begin{widetext}
\beq
\label{5.1}
f_{i,\ell}(\mathbf{r},\mathbf{V};t)=n_i(\mathbf{r};t)v_\text{th}^{-d}(T(\mathbf{r};t))\varphi_i \Bigg(\frac{\mathbf{V}}{v_\text{th}(T(\mathbf{r};t))},\frac{\Delta_{ij}}{v_\text{th}(T(\mathbf{r};t))}\Bigg),
\eeq
\end{widetext}
where $\mathbf{V}=\mathbf{v}-\mathbf{U}(\mathbf{r};t)$ is the peculiar velocity and $\mathbf{U}(\mathbf{r};t)$ is the mean flow velocity of the mixture. On the other hand, given the technical difficulties associated with the Enskog equation, we plan as a first step to consider dilute granular mixtures described by the Boltzmann kinetic equation. In contrast to the results obtained from the IHS model for low-density gases~\cite{GD02,GMD06,GM07}, an interesting feature of the $\Delta$-model is that there will be nonvanishing first-order contributions to the partial temperatures and the cooling rate, which are proportional to $\nabla\cdot \mathbf{U}$. The computation of these contributions along with the Navier--Stokes transport coefficients associated with the mass, momentum, and heat fluxes is in progress and will reported in the near future.

\acknowledgments

The work of R.B.~has been supported by the Spanish Ministerio de Economía y Competitividad through Grant No.~FIS2017-83709-R. The research of R.S.~has been supported by the Fondecyt Grant No.1180791 of ANID (Chile).  The work of V.G.~has been supported by the Spanish Ministerio de Economía y Competitividad through Grant No.~FIS2016-76359-P and by the Junta de Extremadura (Spain) Grant Nos. IB16013 and GR18079, partially financed by ``Fondo Europeo de Desarrollo Regional'' funds.


\appendix
\section{Evaluation of the cooling rate}
\label{appA}


In this Appendix the cooling rates $\zeta_i^*=\sum_i \zeta_{ij}^*$ defined by Eq.\ (\ref{2.16}) are evaluated by using the Maxwellian
approximation \eqref{3.1}. Henceforth, it is understood that the dimensionless quantities of Sec.\ \ref{sec3} will be used and the asterisk will be deleted to simplify the notation. To compute all the integrals, we use the property \eqref{1.9b}, which in the homogeneous state reads
\begin{eqnarray}
& & \int d{\bf c}_{1}h({\bf c}_{1})J_{ij}\left[
{\bf c}_{1}|\varphi_{i},\varphi_{j}\right] =
x_j\chi_{ij}
\left(\frac{\sigma_{ij}}{\overline{\sigma}}\right)^{d-1} \nonumber\\
& & \times \int
\,d{\bf c}_{1}\,\int \,d {\bf c}_{2}\;\varphi_{i}({\bf c}_{1})\varphi_{j}({\bf c}_{2})
\int d\widehat{\boldsymbol {\sigma }}\,\Theta
(\widehat{\boldsymbol {\sigma}} \cdot {\bf
g})(\widehat{\boldsymbol {\sigma }}\cdot {\bf
g})\nonumber \\
&& \times \,\left[ h( {\bf c}_{1}')-h({\bf c}_{1})\right] \;.  \label{a1}
\end{eqnarray}
As in previous calculations \cite{GBS18}, the integral defining the quantities $\zeta_{ij}$ can be divided in two parts; one of them already computed in the conventional IHS model (with $\Delta_{ij}=0$) and the other part involving terms proportional to the parameter $\Delta_{ij}$. Thus,
\beq
\label{a2}
\zeta_{ij}=\zeta_{ij}^{(0)}+\zeta_{ij}^{(1)},
\eeq
where the contribution $\zeta_{ij}^{(0)}$ is \cite{GD99b,G19}
\beqa
\label{a3}
\zeta_{ij}^{(0)}&=&\frac{4\pi^{(d-1)/2}}{d\Gamma\left(\frac{d}{2}\right)}
x_j\chi_{ij}
\left(\frac{\sigma_{ij}}{\overline{\sigma}}\right)^{d-1}\mu_{ji}(1+\al_{ij})\theta_i^{-1/2}\nonumber\\
& & \times
\left(1+\theta_{ij}\right)^{1/2}
\Big[1-\frac{1}{2}\mu_{ji}(1+\alpha_{ij})(1+\theta_{ij})\Big]. \nonumber\\
\eeqa
Let us consider the new contribution $\zeta_{ij}^{(1)}$. It is given by
\begin{widetext}
\beqa
\label{a4}
\zeta_{ij}^{(1)}&=&-\frac{8}{d}\theta_ix_j\chi_{ij}
\left(\frac{\sigma_{ij}}{\overline{\sigma}}\right)^{d-1}\mu_{ji}\Delta_{ij}\int
d{\bf c}_{1}\int d{\bf c}_{2}\varphi_{i}({\bf c}_{1})\varphi_{j}({\bf c}_{2})
\nonumber\\
& & \times \int d\widehat{\boldsymbol {\sigma }}\Theta
(\widehat{\boldsymbol {\sigma}} \cdot {\bf g})(\widehat{\boldsymbol {\sigma}}\cdot {\bf g})\Big[\mu_{ji}\Delta_{ij}
-(\widehat{\boldsymbol {\sigma}}\cdot \mathbf{c}_1)+\mu_{ji}(1+\al_{ij})(\widehat{\boldsymbol {\sigma}}\cdot {\bf g})\Big]\nonumber\\
&=&-\frac{8}{d}\theta_ix_j\chi_{ij}
\left(\frac{\sigma_{ij}}{\overline{\sigma}}\right)^{d-1}\mu_{ji}\Delta_{ij}\int
d{\bf c}_{1}\int d {\bf c}_{2}\varphi_{i}({\bf c}_{1})\varphi_{j}({\bf c}_{2})\Big[B_1 \mu_{ji} \Delta_{ij}g-B_2 (\mathbf{g}\cdot \mathbf{c}_1)+B_2\mu_{ji}(1+\al_{ij})g^2\Big],
\nonumber\\
\eeqa
\end{widetext}
where \cite{NE98}
\beq
\label{a5}
B_k\equiv \int\; d\widehat{\boldsymbol{\sigma}}\,
\Theta (\widehat{{\boldsymbol {\sigma }}}\cdot \mathbf{g})(\widehat{\boldsymbol {\sigma }}\cdot \widehat{\mathbf{g}})^k=\pi^{(d-1)/2} \frac{\Gamma\left(\frac{k+1}{2}\right)}{\Gamma\left(\frac{k+d}{2}\right)}.
\eeq
In order to evaluate $\zeta_{ij}^{(1)}$, one replaces $\varphi_i$ by its Maxwellian approximation \eqref{3.1}
and the result is
\beq
\label{a7}
\zeta_{ij}^{(1)}=-\frac{8}{d}\theta_ix_j\chi_{ij}
\left(\frac{\sigma_{ij}}{\overline{\sigma}}\right)^{d-1}\mu_{ji}\Delta_{ij}(\theta_i\theta_j)^{d/2}
I_\zeta,
\eeq
where
\beqa
\label{a8}
I_\zeta &\equiv& \pi^{-d}\int
\,d{\bf c}_{1}\,\int \,d {\bf c}_{2}\;e^{-\theta_i c_1^2-\theta_j c_2^2}\Big[B_1 \mu_{ji} \Delta_{ij}g \nonumber\\
& &
-B_2 (\mathbf{g}\cdot \mathbf{c}_1)+B_2\mu_{ji}(1+\al_{ij})g^2\Big].
\eeqa
The integral (\ref{a8}) can be performed by the change of
variables
\begin{equation}
{\bf x}={\bf c}_{1}-{\bf c}_{2},\quad {\bf y}=\theta_{i} {\bf c}_{1}+\theta_{j}{\bf c}_{2},
\label{a10}
\end{equation}
with the Jacobian $\left( \theta_{i}+\theta_{j}\right)^{-d}$. With the change of variables \eqref{a10}, the integral $I_\zeta$ is given by
\beqa
\label{a13}
I_\zeta&=&\frac{\pi^{d/2}}{2\Gamma\left(\frac{d}{2}\right)}(\theta_i\theta_j)^{-\frac{d}{2}}\Bigg[
\frac{2\mu_{ji}\Delta_{ij}}{\sqrt{\pi}}\left(\frac{\theta_i+\theta_j}{\theta_i\theta_j}\right)^{1/2}
-\theta_i^{-1}\nonumber\\
& &
+\mu_{ji}(1+\al_{ij})\frac{\theta_i+\theta_j}{\theta_i\theta_j}\Bigg],
\eeqa
where $\Omega_d=2\pi^{d/2}/\Gamma(\frac{d}{2})$ is the total solid angle in $d$ dimensions. With this result,  $\zeta_{ij}^{(1)}$ can be finally written as
\beqa
\label{a14}
\zeta_{ij}^{(1)}&=&-\frac{4\pi^{d/2}}{d\Gamma\left(\frac{d}{2}\right)}x_j\chi_{ij}
\left(\frac{\sigma_{ij}}{\overline{\sigma}}\right)^{d-1}\mu_{ji}\Delta_{ij}\Bigg[
\frac{2\mu_{ji}\Delta_{ij}}{\sqrt{\pi}}\theta_i^{1/2}\nonumber\\
& & \times \left(1+\theta_{ij}\right)^{1/2}
-1+\mu_{ji}(1+\al_{ij})\left(1+\theta_{ij}\right)\Bigg].\nonumber\\
\eeqa
The expression \eqref{3.3} for $\zeta_i$ is easily obtained from Eqs.~\eqref{a3} and \eqref{a14}.



%

\end{document}